\documentclass[12pt]{article}

\usepackage{scicite}
\usepackage{graphicx}
\usepackage{times}
\usepackage{color}

\newenvironment{sciabstract}{%
\begin{quote} \bf}
{\end{quote}}

\newcounter{lastnote}
\newenvironment{scilastnote}{%
\setcounter{lastnote}{\value{enumiv}}%
\addtocounter{lastnote}{+1}%
\begin{list}%
{\arabic{lastnote}.}
{\setlength{\leftmargin}{.22in}}
{\setlength{\labelsep}{.5em}}}
{\end{list}}

\title{ The systematic structure and predictability of urban business diversity}

\author
{Hyejin Youn,$^{1,2,3\ast}$, Lu\'is M. A. Bettencourt,$^{3}$ Jos\'e Lobo,$^{4}$ \\
Deborah Strumsky,$^{5}$ Horacio Samaniego,$^{6}$ and Geoffrey B. West,$^{3}$ 
\\
\normalsize{$^{1}$ Institute for New Economic Thinking, Walton Well Rd, OX2 6ED, Oxford,}\\
\normalsize{$^{2}$ Mathematical Institute, University of Oxford, Woodstock Road, Oxford, OX2 6GG UK, }\\
\normalsize{$^{3}$ Santa Fe Institute, 1399 Hyde Park Road, Santa Fe, NM 87501 USA,}\\
\normalsize{$^{4}$ School of Sustainability, Arizona State University, Tempe, AZ 85287 USA,}\\
\normalsize{$^{5}$ Geography and Earth Sciences, University of North Carolina at Charlotte,}\\
\normalsize{ Charlotte NC 28223 USA,}\\ 
\normalsize{$^{6}$ Facultad de Ciencias Forestales and Recursos Naturales,}\\
\normalsize{Universidad Austral de Chile, Valdivia, Chile}\\
\normalsize{$^\ast$To whom correspondence should be addressed; E-mail: youn@maths.ox.ac.uk}
}

\date{}
\begin{document}

\baselineskip24pt

\maketitle

\begin{sciabstract}
Understanding cities is central to addressing major global challenges from climate and health to economic resilience. Although increasingly perceived as fundamental socio-economic units, the detailed fabric of urban economic activities is only now accessible to comprehensive analyses with the availability of large datasets. Here, we study abundances of business categories across U.S. metropolitan statistical areas to investigate how diversity of economic activities depends on city size.  A universal structure common to all cities is revealed, manifesting self-similarity in internal economic structure as well as aggregated metrics (GDP, patents, crime). A derivation is presented that explains universality and the observed empirical distribution. The model  incorporates a generalized preferential attachment process with ceaseless introduction of new business types. Combined with scaling analyses for individual categories, the theory quantitatively predicts how individual business types systematically change rank with city size, thereby providing a quantitative means for estimating their expected abundances as a function of city size.  These results shed light on processes of economic differentiation with scale, suggesting a general structure for the growth of national economies as integrated urban systems.
\end{sciabstract}

Diversity is central to the resilience of complex adaptive systems whether ecosystems or economies~\cite{Whittaker1972, Quigley1998, PugaGillesDuranton2000}. In particular, it has been argued that the success and resilience of cities, together with their role in innovation and wealth creation, are driven by their ever-expanding diversity~\cite{Quigley1998, PugaGillesDuranton2000, Glaeser1992, Henderson1988, Henderson1995, Henderson1974, Glaeser2011, Jacobs1984}. The presence and ever-changing admixture of individuals, ethnicities, cultural activities, businesses, services, and social interactions is a defining characteristic of urban life. Together with its counterpart, specialization, this is often cited as what to makes a city unique and distinctive, and has consequently featured prominently in the study of cities across economics, geography and urban planning.  Despite its acknowledged importance, however, there have been surprisingly few quantitative investigations into possible systematic regularities and underlying dynamics that govern the diversity of cities across an entire urban system.  A recurrent goal in developing an overarching science of cities is to discover, and conceptually understand, general patterns for how people, infrastructure and economic activity are organized and inter-related\cite{Batty2008, Bettencourt2010b}. A compelling question therefore is how is diversity related to aggregate urban socio-economic and infrastructural metrics and how do these depend on city size\cite{Bettencourt2010a, Bettencourt2007, Bettencourt2012, GomezLievano2012}?  

The systematic quantitative understanding of diversity requries two issues be addressed.  Measuring diversity typically involves identifying different (business) types and counting their frequency for a given unit of analysis, such as a city or a nation~\cite{Whittaker1972}. It should be immediately clear, then, that such a task can be problematic because any systematic classification scheme is subject to an arbitrary recognition of specific categories, since any business type can be subdivided further as long as a defining distinction is made. Restaurants, for example, can be decomposed into fine dining, fast food, etc, and into their cuisine, price, quality etc. In general, therefore, urban diversity is scale-dependent and we should seek a resolution-independent characterization. Secondly, we need to deconvolute the intricate relations between scale, diversity and economic productivity, and between diversification and specialization.  New business types must involve a larger number of people, both as workers and clients, and, to be sustained, should lead to greater economic productivity, by permitting, for example, greater specialization and interdependence. Thus we may expect that the larger scale of bigger cities should afford greater economic diversity (at least in absolute terms) but such an expectation is at odds with the idea that specialization drives increase in efficiency.

In this paper we present a novel approach to measuring and characterizing economic diversity in order to clarify its underlying role in urban economic development. Our analysis reveals a surprising systematic behaviour common to all cities. We show how this can be derived theoretically and present a simple model for understanding its structure based on a variant of preferential attachment for introducing new business types. The model quantitatively predicts how individual business types systematically change rank with city size, shedding light on processes of economic differentiation with scale.

We focus on the frequency distribution of business types (the number of ``species'') 
and first ask how this varies across cities (the ``ecosystem''). We identify our unit of 
analysis as the {\it establishment}, which is defined as a single physical location 
where business is conducted, so that, for example, individual stores of a national 
chain would be counted separately. Establishments are nowadays seen as 
fundamental units of economic analysis because innovation, wealth generation, 
entrepreneurship and job creation all manifest themselves through the formation and 
growth of workplaces~\cite{Haltiwanger1998}.  We explore a unique dataset,
the National Establishment Time-Series, a proprietary longitudinal database built 
by Walls \& Associates, to capture economic life at an extraordinarily 
fine-grained level~\cite{data, hypothesis2013}.  This dataset includes 
records of nearly the entire set of establishments (work places) in US urban 
areas (over 32 million) each of which is classified according to the North American 
Industry Classification System (NAICS).  We aggregate such information 
into the standard definition of functional cities: the 366 Metropolitan Statistical 
Areas (MSAs), which are defined by the census bureau as unified labor markets centered 
on a single large city wielding substantial influence over its surrounding region\cite{MSA}.
These MSAs account for over 90\% of the economic output of the US and almost 85\% of its 
population (the lower limit on MSA size is $\sim 50,000)$. 

The data reveals that the total number of establishments, $N_f$, in each MSA is linearly 
proportional to its population size, $N$: that is, 
\begin{eqnarray}
N_f \approx \eta N
\label{scaling_establishment}
\end{eqnarray}
with the proportionality constant $\eta \simeq 21.6^{-1}$ (Fig 1A).  
Thus, on average, there are about 22 people per establishment in a city, 
regardless of its size. Combined with the fact that the number of employees, 
$N_e$, also scales linearly with $N$, the average size of establishments is 
also independent of population size with $N_e/N_f \simeq 11.9$. This remarkable constancy 
of the average employment rate and number of establishments across cities is puzzling 
when viewed in light of agglomeration effects, and {\it per capita} increases in productivity,
 wages, GDP, or patent production, with population size~\cite{Bettencourt2007, Bettencourt2012}.
Clearly, then, the increasing returns to scale characterizing the benefits of urbanization and increasing city size are not simply due to bigger cities having more establishments. This suggests that investigating the diverse composition of different economic types is an important 
ingredient for understanding how larger cities realize gains in their aggregate economies.

The simplest measure of the level of economic diversity counts distinct establishment types, 
$D(N)$, within a metropolitan area of population $N$~\cite{Whittaker1972}. 
Figure 1B shows that $D(N)$ increases logarithmically with $N$, but eventually 
levels off.  This saturation, which is also observed in Japanese metropolitan 
areas~\cite{Mori2008}, may be attributed to the failure of the finite NAICS 
classification scheme which for the largest cities cannot fully capture the true 
extent of economic diversity: When means of differentiation among work places is insufficient, 
saturation is inevitable.  For example, the category {\it restaurants} in NAICS could 
be further deconstructed into different cuisines: Thai, Indonesia, Korean, Japanese, 
Chinese, etc. This effect can be traced out by utilizing the hierarchical structure of NAICS.
Fig. 1B (colored circles) shows $D$ for different resolutions, $r$, normalized by maximum numbers $D_{\rm max}(r)$ 
for each level.  The logarithmic increase remains intact while the saturation appears 
at larger and larger city sizes as the resolution becomes finer. This suggests that 
saturation is, in fact, an artifact of insufficient detail in the description of 
business types. We return to this point below.

A more insightful way of assessing economic diversity is to examine the constituent
types of $D(N)$ for individual cities. The abundance of the hundred leading business 
types for a selection of cities is shown in Fig. 2A.   
In New York, the most abundant business type is 
{\it offices of physicians}, followed by {\it offices of lawyers} and {\it restaurants}; 
Phoenix ranks {\it restaurants} first and {\it real estate} second (perhaps not 
surprising in a rapidly growing city); San Jose, which includes Silicon Valley, 
predictably ranks {\it computer programming} second only to {\it restaurants}.  
Indeed, the composition of economic activities in cities has its own distinctive 
characteristics reflecting the individuality of each city. {\it It is therefore all 
the more remarkable that, despite the unique admixture of business types for cities, 
the shape of these distributions is universal; so much so that, with a simple scale 
transformation, their rank abundances collapse to a single unique curve common to 
all cities} (Fig. 2B).  Note that the curve is robust to changes in levels of wealth, 
density, and population size, which vary widely across the U.S.

This universality can be derived from a sum rule for the total number of establishments as follows.  Let $F_i(N)$ be the number of the $i^{th}$ most
abundant business type in a city of size $N$, as shown in Fig. 2A. When summed over all 
ranks, this must add up to the total number of establishments, $N_f(N)$: thus, 
$\sum_{i = 1} ^{D_{max}} F_i(N) = N_f(N) = \eta N$. But, from Eq. (\ref{scaling_establishment}) $N_f(N) = \eta N$, so 
introducing the per capita distribution, $ f_i(N)\equiv F_i(N)/N$, this can be re-expressed as 
\begin{eqnarray}
\sum_{i = 1}^{D_{\rm max}} f_i(N) = \eta
\label{eq:sumrule}
\end{eqnarray}
As $N$ increases, any growing or diverging dependences of the $ f_i(N)$ on $N$ in Eq. (\ref{eq:sumrule})
cannot be cancelled against each other because of their positive definiteness, whilst any decreasing dependence vanishes.  Since $\eta$ is a constant this implies that each $f_i$ must itself become independent of $N$ for sufficiently large $N$, predicting that the {\it per capita} frequency of business types must be the same for all cites. Note that the derivation is independent of the underlying dynamics. Fig. 2B verifies the predicted invariance of $f_i$ across all cities.  The invariance remains valid when we treat the discrete rank, $i$, as a continuous variable, $x$, and correspondingly $f_i$ as a continuous function, $f(x)$.  The surprise in the data is that this predicted collapse to a single curve extends all the way down to relatively small cities (that is, up to relatively high ranks) mirroring a similar precocious scaling observed in urban metrics.

The universal form of this scaled rank-size distribution, $f(x)$, has three distinct 
regimes: for small $x (< x_0$, say), it is well described by a Zipfian power law 
with exponent $\gamma$, as shown in the inset of Fig. 2B; for larger $x (> x_0$), 
it is approximately exponential; and finally, as $x$ approaches the maximally 
allowed value for the total number of categories, $D_{\rm max}$, 
$f(x)$ drops off suddenly.  To a very good approximation, these can be combined 
into a single analytic form:
\begin{eqnarray}
f(x) = A x^{-\gamma} e^{-x/x_0} \phi(x, D_{\rm max}).
\label{eq:universalcurve}
\end{eqnarray}
The parameters are $\gamma\approx 0.49$ and $x_0\approx211$.  
The overall normalization, $A$, is not an independent parameter and is determined from 
the sum rule for $f(x)$,  Eq. (\ref{eq:sumrule}),  which gives $A\approx f(1) \approx 0.0019$ in excellent 
agreement with the empirical data.   

The function, $\phi (x, D_{\rm max})$, parametrizes the cut-off that is enforced by the 
finite resolution, $D_{\rm max}$.  It satisfies two conditions: (i) it is only 
important when $x$ approaches $D_{\rm max}$, and completely dominates when 
$x = D_{\rm max}$; (ii) its effect vanishes if the resolution is sufficiently fine.  A simple phenomenological function that satisfies 
all of these conditions is $\phi(x, D_{\rm max}) = e^{[1 - (1 - x/D_{\rm max})^{-1}}]$. 
For comparison, we show fits to the data both with and 
without $\phi$ in Fig 1B and Fig 2B to illustrate the effect of $\phi$.

Recall that $D_{\rm max}(r)$ is the maximum possible number of business categories 
that can appear in a city using a given classification scheme with resolution $r$, 
and therefore increases with increasing resolution. In the limit of the finest grained resolution, $r\rightarrow \infty$, we expect $D_{\max} \to \infty$, in which case the term $\phi$ becomes constant, turning off the saturation phenomenon. This effect is already visible in Fig. 1B, where saturation sets in at smaller city sizes for coarser resolution.  This leads to the conclusion that the highest present level of resolution of the NAICS classification scheme may be insufficient to capture the actual business type diversity of the largest cities.  

The cut-off, $\phi$, is, therefore, attributed to the saturation observed in Fig 1B. 
This can be shown by deriving the diversity function $D(N)$ from the rank-size 
distribution $f(x)$ (including the cut-off $\phi$). This construction is possible because 
$D(N)$ corresponds to the lowest possible rank, $x$, containing only a single establishment. In the continuum limit this gives $F[D(N)] = 1$, or $f[D(N)] = 1/N$, leading to 
\begin{eqnarray}
D(N) = f^{-1}(1/N)
\end{eqnarray} 
Solving this numerically using Eq.~(\ref{eq:universalcurve}) gives excellent agreement with data in Fig. 1B (the black solid line).  When the classification resolution $r$ is sufficiently fine, $D_{\rm max}$ becomes large and $\phi$ becomes 1. 
In such a limit, the numerical solution gives no saturation in $D(N)$ as the orange dashed curve in Fig 1B shows.
The analytic solution for infinite resolution reduces to 

\begin{eqnarray}
D(N) \approx x_0 \ln N/\mathcal{N}
\label{eq:diversity}
\end{eqnarray} 
where $\mathcal{N}$ depends weakly on $N$.  This represents an open-ended ever-expanding diversity with population growth and confirms that the cut-off, $\phi$, is associated with the saturation observed in $D(N)$. 

We can understand the structure of $f(x)$ and $D(N)$ in the context of generalized preferential attachment or growing models \cite{Barabasi1999a, Newman2005}. The growing model is a widely accepted mechanism for generating rank-size distributions, whether for words, genes or cities. It is based on a stochastic growth process in which new elements of the system (business types in this case) are attributed a probability, $\alpha$, of adding a new type, or adding to an existing type~\cite{Yule1925,Simon1955}.  In the classic Simon-Yule model, for example, the attachment probability, $\alpha$, of being an existing type is proportional to the existing frequency.  As a result, the model exhibits a feed-back mechanism in which more-frequent types acquire new elements with higher probability than less-frequent types.  Such a model provides a plausible mechanism for the observed Zipfian behavior in the high frequency regime of $f(x)$: namely, a power-law distribution $x^{-\gamma}$ with $\gamma=(1-\alpha) < 1$ when $x < x_0$ as in Fig 2B. Since $\gamma$ is estimated as 0.49 in the Eq. (\ref{eq:universalcurve}) from our empirical data, $\alpha$ is $0.51$.  This number suggests quite a rapid expansion so that small cities go through a stage of significant increase in business diversity as they grow: One in two new establishments entering the system, on average, spawns a new type.  This fast pace of diversification contrasts to the pace in the exponential regime of $f(x)$ where $x$ is larger than $x_0$. The Yule-Simon model can be still used to explain this exponential regime by relaxing the condition for $\alpha$: The probability is no longer constant but a function of system size $N$.  In our case, the probability of each business being a new type decreases with city size.  In fact, this slow diversification is observed in our Fig 1B where the number of business types $D(N)$ logarithmically increases, that is, slower than the linear.  Because $\alpha (N) = d D(N)/d N$, the logarithmic function $D(N)$ in the Eq. (\ref{eq:diversity}) gives the $\alpha (N) \approx x_0 /N$.  This variation in $\alpha (N)$ can be used in the model of aggregate growth to predict the exponent $\gamma= \frac{\alpha(N) N}{D(N)} \frac{1}{1-\alpha(N)} = [ ( 1- x_0/{\cal N}) \ln(N)]^{-1} $, which vanishes for large $N$, and results in the observed exponential behavior in $f(x)$.

The empirical findings (Figs.~1-2) coupled with the predictions of the model described above suggest that all 
cities, as they grow, exhibit the same underlying dynamics in the development of 
their business ecology.  Initially, small cities, with a limited portfolio of 
economic activities, need to create new functionalities at a fast pace. These basic 
activities constitute the economic core of every city, big and small.  Later, 
as cities grow, the pace at which new functionalities are introduced slows down 
dramatically, but never completely ceases.  Large cities, then, presumably rely primarily on combinatorial processes for developing new relationships among their many existing functionalities, which in turn is the source of observed increases in economic productivity. This is a general feature of combinatorial growth process: Once the set of individual building blocks is large enough, their combination is sufficient to generate novelty even when the set itself expands slowly or not at all.

The universal distribution of frequencies does not, however, account for the entire 
developmental process of economic functionalities in cities.  The stochastic 
Simon-Yule model, for example, does not predict what business compositions sit in what ranks.  
If, during  growth, the introduction and success of each establishment were 
independent of business type (but dependent on frequencies), there would be 
no structure in how ranks  are occupied. This is in clear disagreement with 
the following observation as well as with the pattern that ``creative'' activities and innovation concentrates disproportionally in large cities ~\cite{Bettencourt2007, Henderson1988}.

The process by which specific business types assume different ranks in different cities 
may be particular to the ecology of specific places; or it may also be a 
property of scale.  To distinguish between these two cases, we perform a 
multi-dimensional allometric scaling analysis of the number of specific establishments 
in each type.  The super (or sub)-linearity of specific business types represents a 
systematic {\it per capita} increase (or decrease) of their abundances with city size.
Fig. 3A shows an example: the number of {\it lawyers' offices}, $N_{lo}$, scales as 
$ N_{lo} \sim N^{\beta}$, with $\beta \approx 1.17$. That the exponent $\beta$ is 
greater than one (super-linear) means  that larger cities systematically have more lawyers {\it per capita}.  
Because lawyers' offices typically appear at high frequencies ($x < x_0$) we can approximate 
$f(x)$ in Eq.~(\ref{eq:universalcurve}) by its power law behavior and write,
$N_{lo}/N = f(x_{lo}) \sim x_{lo}^{-\gamma}$ and thereby derive 
how the rank of lawyers changes with city size: $x_{lo} \sim N^{{(1-\beta)}/{\gamma}}$. 
This predicts $x_{lo} \propto N^{-0.4}$, which is in good agreement with the actual 
scaling shown in Fig. 3B. 
We can similarly predict how the ranks of low abundance business types scale. The 
rank shift can be expressed as: 
	\begin{eqnarray}
		x \sim \left\{ \begin {array} {rl}
		N^{\frac{1-\beta}{\gamma}} &\mbox{ for small } x < x_0 \\
		x_0 (1 - \beta) \ln N &\mbox{ for large } x > x_0
				\end{array} \right.  \label{eq:rankchange}
	\end {eqnarray}
Thus, business types whose abundances scale super-linearly with population size 
systematically increase their rankings, whereas those that are sub-linear systematically 
decrease, as expected.

Fig.~3C summarizes the values of scaling exponents for business sectors at the 2-digit-level. 
Most primary sectors such as, {\it agriculture}, {\it mining}, and {\it utility} scale sub-linearly, predicting their systematic 
suppression, in relative terms, as cities get larger.  On the other hand, informational 
and service businesses, such as {\it professional, scientific, and technical services} 
and {\it management of companies and enterprises} scale super-linearly, and are 
therefore predicted to increase disproportionally with city size, as observed. There are also 
sectors such as {\it restaurants}, for example, that do not change ranks.  Note that sectors that deviate from 
linearity tend to be tradable industries that may be exchanged across cities~\cite{Sachs1994}.  
Because markets for these industries are not restricted to their immediate spatial location, 
comparative advantages may generate agglomeration effects resulting from city size and/or 
to specific places.


We have shown that the distribution of business in U.S. cities is characterized by a universal rank-size curve in which specific types predictably increase or decrease their relative rankings and frequencies as a function of city size.  The results constitute a first general picture of the properties of the economic diversity of US cities measured in terms of business types. We show how statistics of business composition may be a result of a general mechanism of business creation based on existing frequencies and how the rate of introduction of new types slows down with city size.  Our results, together with the theoretical framework presented, provides an important contribution to developing a theory of cities and urbanization that can encompass their dynamics, organization, and economic diversity. It provides  quantitative support for the hypothesis advanced by classic central place and locational theories of cities~\cite{Christaller1966, Losch1954} that a general hierarchy of economic activities exists. These classic observations need to be interpreted not only in terms of the appearance of specific new sectors with city size, but also with the {\it disproportionate} growth of {\it certain} types versus others. Our work also gives quantitative support to business life-cycle theories~\cite{Klepper1997,Duranton_Puga2001}, where some types of business may be more prevalent in larger cities, but over time tend to move down the urban hierarchy as they mature and internalize more of their business model. We believe that the present results, together with further analyses of revenue, employment, temporal patterns, provide the foundation for a mechanistic understanding of how large cities realize greater economic productivity and how urbanization tends to promote nationwide economic growth~\cite{Bettencourt2012b}.

\section{Materials and Methods}

The National Establishment Time-Series is a proprietary longitudinal database built by Walls \& Associates. This dataset includes records of nearly the entire set of {\it establishments} in US urban areas (over 32 million) each of which is classified according to the North American Industry Classification System (NAICS). The establishment is a term to define a single physical location where the business is conducted (work place). Therefore one city can have multiple number of Starbucks establishments. We aggregate these establishments of various NAICS categories within a city. We use Metropolitan Statistical Area (MSA) as a functional city for our unit of analysis.

\begin{scilastnote}
\item We thank Ricardo Hausmann, Coco Krumme and Marcus Hamilton for discussions. This research is partially supported by the Rockefeller Foundation, the James S. McDonnell Foundation (grant no. 220020195), the National Science Foundation (grant no. 103522), the John Templeton Foundation (grant no. 15705), the U.S. Department of Energy through the LANL/LDRD Program (contract no. DE-AC52-06NA25396), and by gifts from the Bryan J. and June B. Zwan Foundation and the Eugene and Clare Thaw Charitable Trust.
\end{scilastnote}

\begin{figure}
	\centerline{\includegraphics[width=0.45\textwidth]{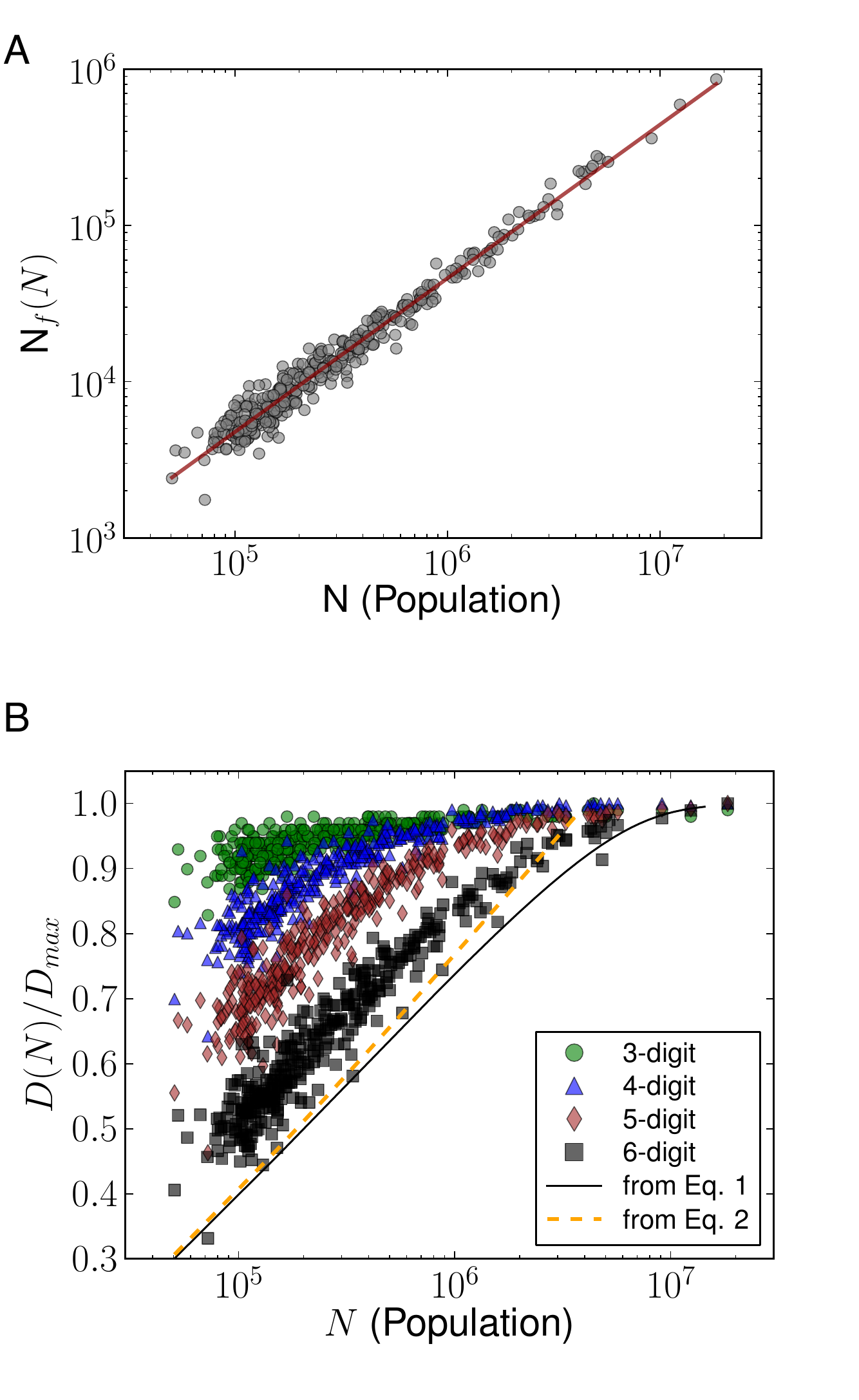}}
	\caption{{\bf The total number of establishments and their unique types as a function of city size}.
	{\bf (A)} The total number of firms $N_f$ scales linearly with city size:
	$N_f \sim N^{\alpha}$ where $\alpha=0.98 \pm 0.02$ with $R^2=0.97$. {\bf (B)}
	The number of distinct business types $D$ normalized by its maximum value
	$D_{\rm max}$ at various levels of classification, $r$, based on the NAICS scheme, from the
	lowest resolution (3-digit) to the highest (6-digit) denoted by green circles,
	blue triangles, red  diamonds and black squares, respectively (corresponding values of
	$D_{\rm max}$ are 317, 722 and 1160).  All values are scaled by the corresponding
	size of the classification scheme at that resolution, $D_{\rm max}$, such that
	all values fall in between 0 and 1.  Note that $D (N)$ behaves very similarily to 
	data of Japan~\cite{Mori2008}. The black solid line and orange dashes are the predictions 
	from Eq. 1. with and without $\phi$. 
} \label{fig:scaling_diversity}
\end{figure}

\begin{figure*}
  \centerline{\includegraphics[width=0.99\textwidth]{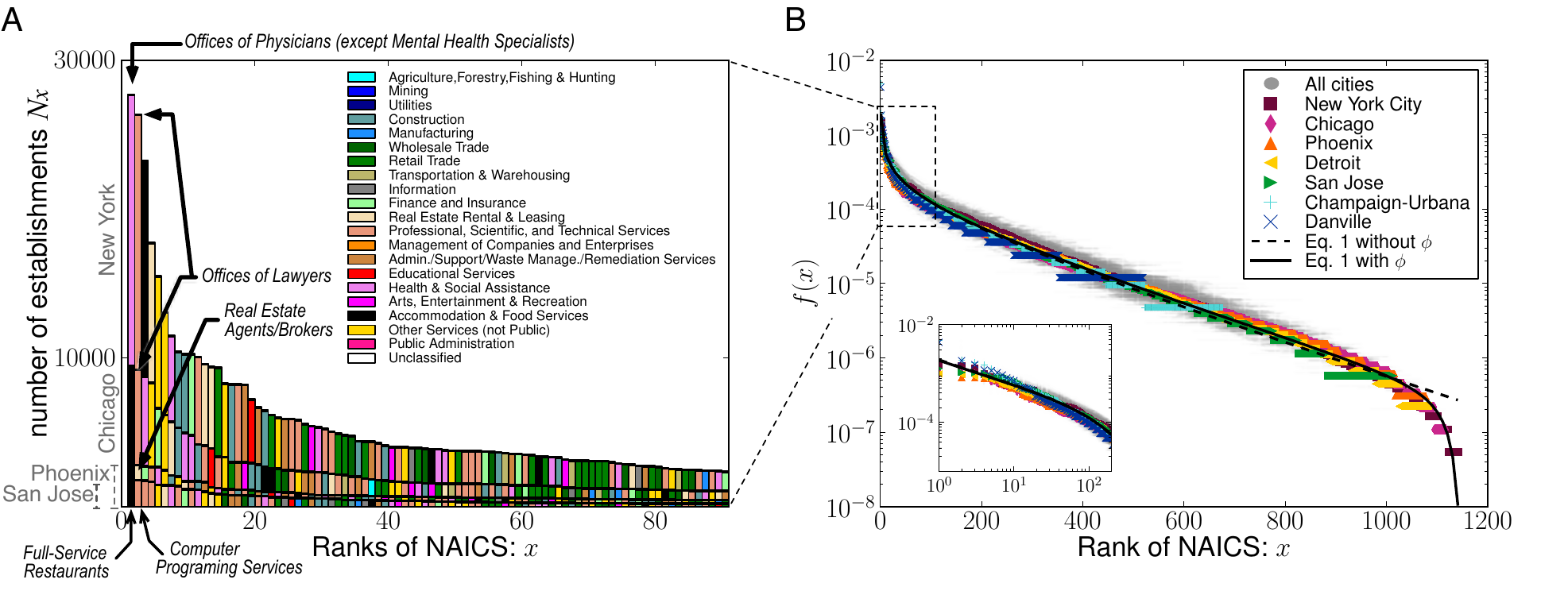}} 
  \caption{{\bf Rank-abundance of establishment types} (A) The number
	of establishments at rank $x$ ranging from 1 to 90 in descending order of their
	frequencies (from common to rare) for New York city, Chicago, Phoenix and San Jose.
	Establishment types are color coded by their classification at the 2-digit level.
	(B) Universal rank-abundance shape of the establishment type by dividing
	$N_x$ by the population size of city in semi-log for all ranges. All metropolitan
	statistical areas are denoted by gray circles. Seven selected cities are denoted
	by various colors and shapes; New York city, Chicago, Phoenix, Detroit, San Jose, Champaign-Urbana,
	and Danville are, respectively, marked by red squares, pink diamonds, orange triangles,
	yellow left triangles, green right triangles, sky blue pluses, and blue crosses.
	The black dash line and the black sold line are fits predicted from Eq. 1 without and with $\phi$ respectively. The inset shows the first
	200 types on a log-log plot showing an approximate Zipf-like power law behavior.}
  \label{fig:collapse}
\end{figure*}

\begin{figure}
  \centerline{\includegraphics[width=.45\textwidth]{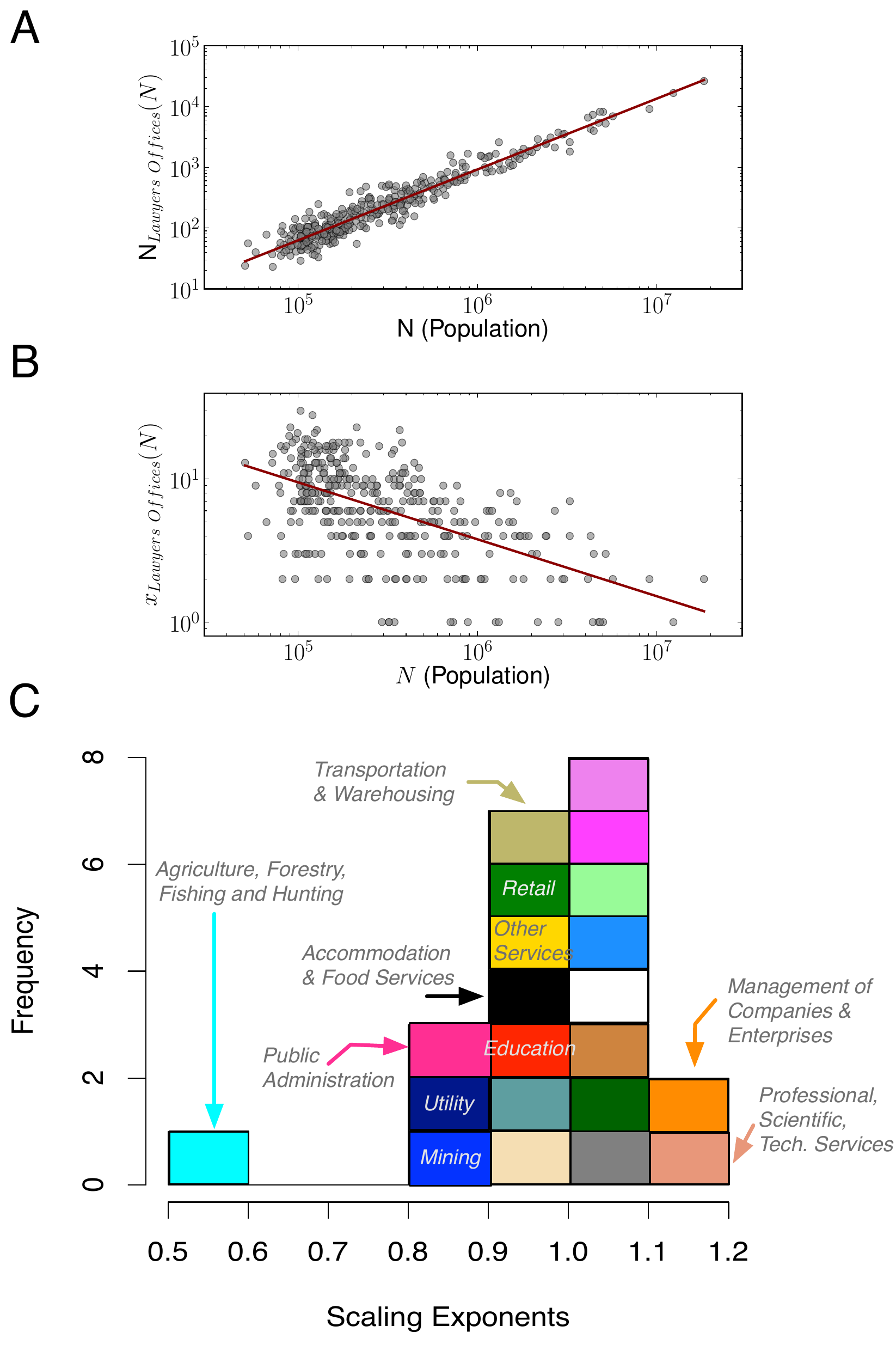}}
  \caption{{\bf Multi-dimensional allometric scaling of industry types}.
	{\bf (A)} The number of lawyers' offices scales superlinearly with population size,
	$N_{\rm lo} \sim N^{1.17 \pm 0.04}$ with $R^2 = 0.92$. {\bf (B)} The rank of lawyers'
	offices increases with  population size, expressing an increase in their relative abundance:
	$x_{\rm lo} \sim N^{-0.4 \pm 0.06}$ with $R^2 = 0.32$. {\bf (C)} Histogram of
	scaling exponents $\gamma$ for all establishment types at the 2-digit level.
	While primary sectors disappear, managerial, professional, technical
	and scientific firms increase in relative abundance, helping to explain
	the increased productivity of larger cities despite the slow addition
	of new business types. }
  \label{fig:scaling_exponents}
\end{figure}

\end{document}